\documentclass[12pt]{article}
\usepackage{graphicx}% Include figure files
\newcommand{\sss}{\scriptscriptstyle}

\newcommand {\be}{\begin{equation}} % start equation
\newcommand{\ee}{\end{equation}}    % end equation

\def\dds{\frac{\partial}{\partial s}}
\def\dds1{\frac{\partial}{\partial s_1}}

\def\vti{v_{{\sss T}i}}
\def\vte{v_{{\sss T}e}}
\def\vtj{v_{{\sss T}j}}

\def\vtse{v_{{\sss T}se}}
\def\vtsi{v_{{\sss T}si}}
\def\vtsj{v_{{\sss T}sj}}
\def\vtfe{v_{{\sss T}fe}}
\def\vtfi{v_{{\sss T}fi}}
\def\vtfj{v_{{\sss T}fj}}
\def\d{d\kern-0.8 ex\vrule height 1.3 ex depth-1.24 ex width 0.7 ex
\kern 0.15 ex}
\def\D{D\kern-1.7 ex\vrule height .87 ex depth-0.8 ex width 0.7 ex
\kern 0.95 ex}

\begin{document}

\baselineskip 18 pt
\begin{center}
  {\Large{\bf New features of   ion acoustic waves in inhomogeneous and permeating  plasmas}}
\vspace{0.5cm}

{\bf   J. Vranjes} \\
      Institute of Physics Belgrade, Pregrevica 118, 11080 Zemun, Serbia\\

\end{center}
 \thispagestyle{empty}

  \abstract
  {  It is generally accepted that the ion acoustic (IA) wave in plasmas containing ions and electrons with the same temperature   is of minor importance due to strong damping of the wave by hot resonant ions.
      In this work it will be  shown  that the IA wave  is  susceptible to excitation even in plasmas with hot ions when both an electromagnetic transverse wave and a background density gradient are present in the plasma, and in addition the wave is in fact unstable (i.e., growing) in the case of permeating homogeneous plasmas.   The multi-component fluid theory is used to describe the IA wave susceptibility for excitation in inhomogeneous plasmas and its coupling with electromagnetic  waves. The growing IA wave in permeating homogeneous plasmas is described by the kinetic theory.
     In plasmas with density and temperature gradients the IA wave is effectively coupled with the electromagnetic waves. In comparison to ordinary IA wave in homogeneous plasma,  the  Landau damping of the present wave is  much smaller, and to demonstrate this effect a simple but accurate fluid model is presented for the Landau damping.  In the case of permeating plasmas, a kinetic mechanism for the current-less IA wave instability  is presented, with a very low  threshold for excitation as compared  with ordinary electron-current-driven kinetic instability. Such growing IA waves can effectively heat plasma in the upper solar atmosphere by a stochastic heating mechanism presented in the work.
  The results of this  work suggest that the IA wave role in the heating of the solar atmosphere (chromosphere and corona) should be reexamined.

}
  % \keywords{Physical data and processes: Waves, Hydrodynamics
%Instabilities--Sun: atmosphere
%                               }

 %  \maketitle
%
%________________________________________________________________

\section{Introduction }\label{s1}

Acoustic waves have been a popular tool in models dealing with the heating in the solar atmosphere
from early days of the solar plasma physics   (Alfv\'{e}n  \cite{alf1}, Biermann \cite{bir1}, Schwarzschild \cite{sw1}, D'Angelo \cite{da1}). However, unsupported by observations, in the course
of time   these models have been replaced by many others based on different  plasma modes and on various different heating mechanisms.
Though, very recent observations (Bello Gonz\'{a}lez  et al.  \cite{belo},  Kneer  and Bello Gonz\'{a}lez  \cite{kn}) have revived interest in the acoustic modes. This because large  fluxes of acoustic modes excited by  convective motion in the photosphere have been observed, that are only by a factor 2 smaller than the value  presently accepted as required for a sustainable heating of the chromosphere.

Models based on wave heating in the solar atmosphere are typically based on a scheme which implies the initiation  of such waves in  lower layers and  their consequent   propagation towards the upper solar atmosphere where they are supposed to dissipate and deposit their energy. Hence, if observed flux  of such a wave is  too low or absent, its  role in the heating is assumed as unimportant. However, wave heating scenario may include models with waves that are generated directly on the spot, i.e., in the corona itself and then dissipated contributing to heating, see examples based on the drift wave theory presented in Vranjes and Poedts (\cite{v2,v3, v4,v5}) and in  Vranjes (\cite{v6}, \cite{v6a}). This implies the presence of some energy source for excitation of such waves directly in the corona. In the case of the mentioned drift wave model the energy is stored in the gradients of background density and temperature.

In the present work we shall show that the ion acoustic waves may be produced on massive scales directly in the upper solar atmosphere. Their energy may  further be transformed into internal energy of the plasma by at least two different mechanisms which will be discussed. We shall  describe two very different possible sources for  their excitation.

 a) The first source are  simply electromagnetic transverse (light) waves that are undoubtedly widespread in  an environment such as the solar atmosphere. It will be shown that  in the presence of such  waves,  ion acoustic waves  are  susceptible to excitation  provided a simultaneous  presence of density gradient. A typical environment for this  mechanism to work are solar magnetic structures where a density gradient should be ubiquitous in the  direction perpendicular to the magnetic field. However, it should be stressed that the magnetic field plays no role in  this excitation mechanism; its role reduces  only to  the equilibrium plasma confinement.  All what is necessary is the presence of a transverse  wave and density gradient. The later can in fact be balanced by a temperature gradient, and the background magnetic field in such cases is not necessary anyhow. An example of this kind will be given in the text.

Frequencies of the ion acoustic (IA) waves from one side   and plasma Langmuir (PL) and electromagnetic (EM, light) waves from the other, are well separated. The former is normally below the ion plasma frequency, while  the Langmuir and EM waves have  a lower cut-off at the {\em electron} plasma frequency. Within linear theory these modes  are  normally uncoupled, yet the situation is quite different in nonlinear regime when IA-PL-EM waves are coupled through so called  parametric decay instability, or IA-EM  waves in the process called  stimulated Brillouin back-scattering (Chen \cite{ch}). Contrary to this, the coupling presented in Sec.~\ref{1}  is essentially  linear.

 b) The second source of the IA wave are inflows of plasmas from some layers to other layers; examples of that kind are numerous and widespread in the  solar
 atmosphere. Such phenomena   create an environment which  can be called permeating plasmas (Vranjes et al. \cite{perm},  Vranjes \cite{v9}), and the presence of these  macroscopic motions implies free energy which can be channeled into excitation of ion acoustic waves. This will be discussed in detail in Sec.~\ref{gro}.

In the solar atmosphere  where the ion and electron temperatures are of the same order, the IA wave is normally expected to be strongly
damped due to kinetic effects. However in the presence of some energy source which can compensate for losses due to the Landau damping, the IA wave can develop in such an environment as well. Typically this can happen
in the case of electron current when the IA wave is growing providing that the equilibrium electron speed exceeds certain limit. For coronal temperatures
on the order of million K this critical electron speed is around $27 c_s$ where $c_s$ is the sound speed. Hence, the instability threshold is rather high.
However, a much lower instability threshold is required for an another type of the kinetic instability in which one plasma propagates through another. Such permeating plasmas are in fact frequently seen in the solar atmosphere in which the plasma from lower layers is accelerated towards upper layers. Those upper layers already contain a plasma (described by some different parameters)  so that we have perfect examples of permeating plasmas throughout the solar atmosphere. This new effect has been discovered in our very recent  works and it has been applied to  cometary plasmas permeated by the solar/stellar winds (Vranjes  \cite{v9}) where it was shown to produce  dust-acoustic  mode that is  practically always growing. In the present work we shall show that the instability is rather effective in the solar atmosphere  as well.

Regardless which of the two suggested  widespread sources may be in action, the excited ion acoustic waves may   be dissipated most effectively by two processes. i) The first one is physically  purely kinetic, the Landau damping mechanism,  when the energy of the wave is transformed first into particle acceleration and then eventually into heat due to collisions. Within  the IA  wave excitation mechanism a) mentioned above we  shall present {\em a fluid model} for its Landau damping. ii) The second process of wave energy dissipation is stochastic and it again implies a  wave-particle interaction which directly leads to stochastic (chaotic) motion and heating.

 Regarding  this dissipation mechanism  ii), it should be stressed   that  most of plasma modes which follow from  multi-component plasma theory are known to cause
stochastic heating. This holds  particularly for electrostatic modes like the ion acoustic wave, Langmuir wave,
lower and upper hybrid waves, ion and electron cyclotron waves, ion and electron Bernstein waves, drift waves (oblique and transverse), and standing waves that can be some of those mentioned here. The electromagnetic modes,  like the Alfv\'{e}n wave, can also heat the plasma by the same mechanism, though as a rule the heating rate is lower and the mechanism
requires large mode amplitudes. More precisely, for  the Alfv\'{e}n wave this implies strongly nonlinear regime with the perturbed to equilibrium magnetic
field ratio $B_1/B_0\geq 0.75$ (Smith and Kaufman \cite{sk1}, \cite{sk2}). Compare this with the requirement for the stochastic heating with the ion acoustic  wave, where the necessary density perturbation threshold is only $n_1/n_0=0.077$ (Smith and Kaufman \cite{sk1}, \cite{sk2}), and the heating is in fact much more efficient.

In terms of the perturbed electric field, for practically all mentioned electrostatic    modes there is a critical electric field amplitude which is needed for the stochastic heating to take place.
  For the oblique drift (OD) wave the stochastic heating takes
place if $ k_{\bot}^2 \rho_i^2 e \phi/(\kappa T_i) \geq 1$. It causes the heating of bulk plasma, it heats better heavier ions than light ions, and it acts mainly in the perpendicular direction. Details of this mechanism in application to the solar corona can be found in Vranjes and Poedts (\cite{v2,v3,v4, v5}). In the case of the transverse drift (TD) wave we have an electromagnetic mode, yet even in this case one obtains the required amplitude of the perturbed electric field  $k_{\bot}^2 E_{z1}^2/(\omega^2 B_0^2)>1$. Here, $E_{z1}$ denotes the electric field which is strictly parallel to the magnetic field, hence we have an effective acceleration of particles that is accompanied with the stochastic heating in both parallel and perpendicular directions, and  with transport in the perpendicular direction. In this case the heating acts on both ions and electrons; more details are available in  Vranjes (\cite{v6}, \cite{v6a}).

For  standing waves the necessary wave electric field is $e k E/(m \omega^2)\approx 0.15-0.25$. Details on this mechanism can be found in Hsu et al. (\cite{hsu}), and in Doveil (\cite{dov}). All these criteria and features of stochastic heating have been verified in numerous  laboratory experiments; some of the mentioned references contain experimental confirmation of the phenomenon.

\section{ Ion acoustic wave in inhomogeneous environment: excitation in the presence of and coupling with the  transverse electromagnetic  (light) waves }\label{1}

\subsection{ Derivation of wave equation }\label{1a}

We shall assume a plasma immersed in an external magnetic field $\vec B_0=B_0\vec e_z$. The magnetic field is here  only in order to have a physically justified equilibrium. In case of  a density gradient in perpendicular direction $\nabla n_0=n_0'(x)\vec e_x$, the equilibrium yields the diamagnetic drift speed $\vec v_{j0}\equiv v_{j0}\vec e_y=\vec e_z\times \nabla p_{j0}/(q_j n_{j0}B_0)=\vtj^2/(L_n \Omega_j)$, $\vtj^2= \kappa T_j/m_j$, $L_n= (n_0'/n_0)^{-1}$.

For ion perturbations with the phase speed $\omega/k$ on the order of ion  thermal speed,
it may be  necessary to include  an equation for ion temperature (energy) which may be written in the form (Weiland \cite{wei}, Vranjes et al. \cite{v08}):
\be
\frac{3 n_i}{2} \frac{\partial T_i}{\partial t} + n_i T_i \nabla\cdot \vec v_i + \frac{3 n_i}{2}\left(\vec v_i\cdot\nabla\right) T_i=0.
\label{en}
\ee
For a constant equilibrium temperature the linearized energy equation reduces to
\be
\frac{3 n_0}{2} \frac{\partial T_{j1}}{\partial t} + n_0 T_0 \nabla\cdot \vec v_{j1} =0.
\label{en}
\ee
We assume quasi-neutrality both in equilibrium and in perturbed state, and $T_{j0}=T_0$. The linearized momentum  and continuity equations
are
\be
m_jn_0\frac{\partial \vec v_{j1}}{\partial t}=q_j n_0 \vec E_1 - \kappa T_0 \nabla n_1 - \kappa T_{j1} \nabla n_0 - \kappa n_0 \nabla T_{j1},\label{me}
\ee
\be
 \frac{\partial n_1}{\partial t} + n_0 \nabla\cdot \vec v_{j1} +  \vec v_{j1} \nabla n_0=0. \label{ce}
 \ee
 In Eq.~(\ref{me}) the magnetic field effect is completely neglected. The explanation for this is as follows. The assumed perturbations are due to both longitudinal and transverse electric field. The former is responsible for compressional perturbations that eventually lead to ion acoustic and Langmuir perturbations.  The latter (transverse) is due to assumed presence of electromagnetic perturbations.  The omitted linearized Lorentz force term yields two terms $\vec v\times \vec B\Rightarrow  \vec v_{\bot 1}\times \vec B_0+ \vec v_0\times \vec B_1$. Here, $ \vec v_{\bot 1}$ is clearly only due to the transverse electric field which has the frequency
  \be
  \omega_{tr}>\omega_{pe}\gg \Omega_j.\label{freq}
  \ee
   This perpendicular movement  changes direction many times within the hypothetical gyro-period.  In other words, the perturbed perpendicular dynamics corresponds to that of an un-magnetized plasma  so that the Lorentz force is negligible and particle motion is almost without any drift features.
  The second contribution   from the Lorentz force $\vec v_0\times \vec B_1$ is also negligible because the ratio of the second and first  terms  $v_0/v_{ph}$, $v_{ph}=\omega_{tr}/k$  is a small quantity.

 Hence, the remaining displacement in the perpendicular $x$ and $y$ directions due to the transverse time-varying electric field is well-known from books. It can easily be calculated yielding known results
  \[
  x=[q_j E_1/(m(\omega_{tr}^2-\Omega_j^2))] [\cos (\Omega_j t) - \cos(\omega_{tr} t)],
  \]
  \[
    y=[q_j E_1/(m(\omega_{tr}^2-\Omega_j^2))][(\Omega_j/\omega_{tr}) [\sin (\omega_{tr} t) - \sin (\Omega_j t)].
    \]
     Making a plot $x(t)$ and $y(t)$ confirms
 the assumed model as long as the condition (\ref{freq}) is satisfied.

In Eq.~(\ref{ce}) the equilibrium diamagnetic drift contribution is exactly zero as long as the magnetic field is constant (Weiland \cite{wei}, Vranjes and Poedts \cite{v09}). Hence, for perturbations of the shape $f(x)\exp(-i \omega t+ i k z)$, from Eqs.~(\ref{me}, \ref{ce}) we have
\be
\vec v_{j1}=\frac{i q_j \vec E_1}{m_j \omega} - \frac{i \vtj^2}{\omega} \frac{\nabla n_1}{n_0} -\frac{i \kappa T_{j1}}{m_j \omega} \frac{\nabla n_0}{n_0}
-
\frac{i \kappa}{m_j \omega} \nabla T_{j1}, \label{e2}
\ee
\be
-i \omega\frac{n_1}{n_0} + \nabla\cdot \vec v_{j1}+  \vec v_{j1} \frac{\nabla n_0}{n_0}=0. \label{e3}
\ee
From Eq.~(\ref{en}) the perturbed temperature is
\be
T_{j1}=-\frac{i2T_0}{3 \omega}  \nabla\cdot \vec v_{j1}. \label{e4}
\ee
Now we calculate $\nabla\cdot \vec v_{j1}$ and use it in Eqs.~(\ref{e3},\ref{e4}). This yields
\be
T_{j1}=\frac{2}{3} \frac{T_0}{\omega^2- 2 k^2 \vtj^2/3} \left(\frac{q_j}{m_j} \nabla\cdot\vec E_1 + k^2 \vtj^2 \frac{n_1}{n_0}\right),
\label{t}
\ee
and we have the number density expressed through the perturbed electric field
\be
n_1=\frac{q_j n_0}{m_j \omega_j^2}\left(1+ \delta_{j2}\right) \nabla\cdot\vec E_1 +  \frac{q_j \vec E_1}{m_j \omega_j^2} \nabla n_0,
\label{e5}
\ee
\[
\omega_j^2\equiv \omega^2 - k^2 \vtj^2 - \delta_{j1}, \quad  \delta_{j1}= \delta_{j3} k^4 \vtj^4, \quad  \delta_{j2}= \delta_{j3} k^2 \vtj^2,
 \]
 \[
 \delta_{j3}=  \frac{2}{3}\frac{1}{\omega^2- 2 k^2 \vtj^2/3}.
\]
In derivations we used the assumption of local approximation and small equilibrium gradients. Thus, we  neglected the second order derivative of the equilibrium density gradient, as well as products of first derivatives, and used the fact that $\vec k\bot \nabla n_0$.
The terms $\delta_{j1, 2, 3}$  originate from temperature perturbations. Hence, setting
them equal to zero (see later in the text) is equivalent to reducing derivations to isothermal case.

The purpose of the derivation performed so far is to calculate the current which appears in the general wave equation
\be
c^2k^2\vec E_1- c^2 \vec k (\vec k\cdot \vec E_1) - \omega^2 \vec E_1 - \frac{i \omega \vec j_1}{\varepsilon_0}=0,
\label{e6}
\ee
\[
 \vec j_1= n_0e \left(\vec v_{i1}-\vec v_{e1}\right).
 \]
Using Eq.~(\ref{e5}) we may express $T_{j1}$ in Eq.~(\ref{t}) through the electric field only. We then calculate $\nabla T_{j1}$ and $\nabla n_1$, and with all this from Eq.~(\ref{e2}) we finally obtain the perturbed velocity expressed through the perturbed electric field and equilibrium density gradient:
\[
\vec v_{j1}\!=\!\frac{i q_j  \vec E_1}{m_j \omega} - \frac{i q_j  \vtj^2}{m_j \omega}\!\left\{\!\frac{1+ \delta_{j2}}{\omega_j^2} + \delta_{j3} \left[\!1+ \frac{k^2\vtj^2}{\omega_j^2} (1+ \delta_{j2})\! \right]\!\right\}\! \nabla \!\left( \nabla\!\cdot\! \vec E_1\!\right)
\]
\[
-\frac{i q_j  \vtj^2}{m_j \omega}\frac{\nabla n_0}{n_0} \left\{\frac{1+ \delta_{j2}}{\omega_j^2} + \delta_{j3} \left[1+ \frac{k^2\vtj^2}{\omega_j^2} (1+ \delta_{j2}) \right] \right\} \nabla\!\cdot\!\vec E_1
\]
\be
-\frac{i q_j  \vtj^2}{m_j \omega}\frac{n_0'}{n_0} \frac{1}{\omega_j^2} \left(1+ \delta_{j3} k^2 \vtj^2\right) \nabla E_{1x}. \label{e7}
\ee
Using this in Eq.~(\ref{e6}) yields the wave equation with the perturbed electric field only.

\subsubsection{Isothermal perturbations and cold ion case }\label{cold}
We are interested to show some  features and effects of coupling between the ion acoustic and electromagnetic transverse (i.e., light) waves. In view of very lengthy expressions, analytically this is most clearly done  in the limit of cold ions and for isothermal electron perturbations. For this purpose we take $\delta_j=0$, and after setting Eq.~(\ref{e7}) into the wave equation (\ref{e6}) we obtain
\[
c^2 k^2 \vec E_1 - c^2 \vec k \left(\vec k\!\cdot \! \vec E_1\right)- \omega^2 \vec E_1 + \left(\!\omega_{pe}^2 \!+ \omega_{pi}^2\!\right) \vec E_1
-\frac{\omega_{pe}^2 \vte^2}{\omega_e^2}\frac{n_0'}{n_0} \nabla E_{1x}
\]
\be
- \frac{\omega_{pe}^2\vte^2}{\omega_e^2} \nabla\left(\nabla\!\cdot \!\vec E_1\right)
-\frac{\omega_{pe}^2 \vte^2}{\omega_e^2} \frac{\nabla n_0}{n_0} \nabla\cdot \vec E_1=0,
 \label{e8}
\ee
\[
 \omega_e^2=\omega^2-k^2\vte^2.
\]
The $y$-component of this vector equation yields one light wave decoupled from the rest $\omega^2= \omega_{pe}^2 + \omega_{pi}^2
+ c^2 k^2$. However, the $z$ and $x$ components yield the longitudinal electrostatic (ion-acoustic plus Langmuir) waves, and yet another light transverse wave, coupled through the equilibrium density gradient
\be
\left( - \omega^2 +  \omega_{pe}^2 + \omega_{pi}^2   + \frac{k^2 \vte^2 \omega_{pe}^2}{\omega_e^2}\right)E_{1z}-\frac{i k \omega_{pe}^2\vte^2}{\omega_e^2}\frac{n_0'}{n_0}  E_{1x}=0,\label{e9}
\ee
\be
\left( - \omega^2 + \omega_{pe}^2 + \omega_{pi}^2  + k^2 c^2\right)E_{1x}- \frac{i k \omega_{pe}^2\vte^2}{\omega_e^2}\frac{n_0'}{n_0}   E_{1z}=0.\label{e10}
\ee
Eq.~(\ref{e9}) describes longitudinal electric field $E_{1z}$ and corresponding wave oscillations coupled  to  the transverse electromagnetic wave $E_{1x}$ in the presence of the given density gradient.

\begin{figure}[!htb]
   \centering
  \includegraphics[height=6cm,bb=16 14 265 208,clip=]{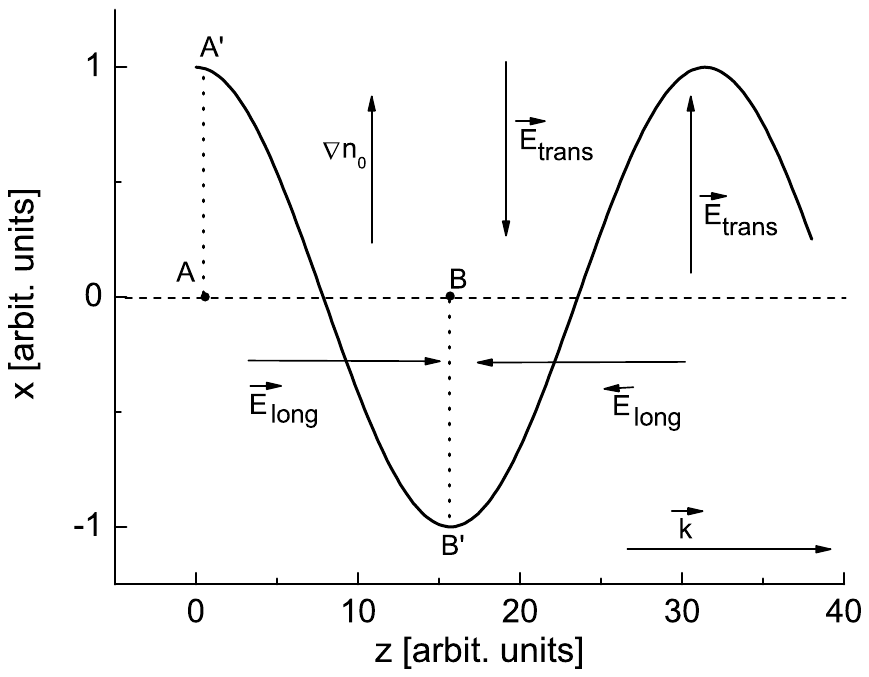}
      \caption{Mechanism of excitation   of longitudinal electrostatic field $\vec E_{long}\equiv\vec E_z$  in the presence of a transverse wave with electric field in $x$ direction which  propagates in $z$-direction.    \label{mech}}
       \end{figure}

 The mechanism of  excitation of the IA wave in the inhomogeneous environment which simultaneously contains EM waves, and its coupling with the latter can be understood by comparing the cases without and with the density gradient, see Fig.~\ref{mech}.
Let us first see  details of  the case {\em with} density gradient in $x$-direction. For this we may focus on one layer with certain density $n_0$ at the position $x_0$. Obviously, this density is initially the same for any point in $z$-direction which has the same position $x_0$. In this layer we may chose two fixed arbitrary points $A(x_0, z_1)$ and $B(x_0, z_2)$ at two different positions in $z$-direction.  Due to the  $E_{1x}$ field the plasma in this layer will be displaced in $x$-direction in such a way as to follow a sinusoidal profile in $z$-direction.  The sinusoidal line in Fig.~\ref{mech} represents displacement of a layer which  was initially at $x=0$, caused by the transverse electric field $\vec E_{trans}\equiv\vec E_x$.  This will have  the following consequences.   i)  The densities $n(x_0, z_1)$, $n(x_0, z_2)$ at the points $A(x_0, z_1)$ and $B(x_0, z_2)$ will no longer be the same. ii) In addition, because of opposite motion of electrons and ions due to the transverse electric field in $x$-direction, there will be excess of charge of one or another sign at every point for any $z$, and this holds for the points $A(x_0, z_1)$ and $B(x_0, z_2)$  as well. The reason for this is clear: when one species (e.g. ions) move from arbitrary point (let it be $A$) along $x$, they are replaced by the particles of the same species which come from some other point in $x$ direction, which has different (e.g. smaller) density. On the other hand electrons from this same point move in opposite direction and and they are replaced by electrons which come from opposite side as compared to moving ions. So there will be more electrons in the point $A$.  Something similar happens at the point $B$.  From i) and ii) it follows that there will be  difference of charge in the chosen arbitrary points  $A(x_0, z_1)$ and $B(x_0, z_2)$. This further implies that there will be  an additional electric field $E_{1z}$ in the $z$ direction.  This  excited longitudinal electric will be accompanied with corresponding density oscillations and this will result in ion acoustic and Langmuir waves. This all is depicted in Fig.~\ref{mech} where $x_0=0$.   In view of the mass difference,  electrons are expected to play main role in the excitation of the longitudinal electric field.

 From the presented model  it is also clear that the equilibrium magnetic field plays no role in the process, as we correctly explained  earlier. This is because the frequency of the transverse oscillations is much higher than the particle gyro-frequency, i.e., the particle is unmagnetized. This is in agreement with standard theory of transverse light waves, which in the high-frequency range predicts the same behavior of the EM wave  in plasmas with or without background  magnetic field (Chen \cite{ch}, Bellan \cite{bel}).  The only essential ingredients are therefore the  transverse oscillating electric field, and the density gradient in the same direction.

 If we now focus on  the case {\em without} density gradient it becomes clear that the effects described above vanish.   There is no electric field in $z$-direction because the density is constant everywhere.

 The process described above will be quantified and confirmed by using realistic parameters for both laboratory and solar plasmas, isothermal and non-isothermal.

The dispersion relation that follows from Eqs.~(\ref{e9}, \ref{e10}) reads
\[
\left(\omega^2- k^2 \vte^2\right) \left[ \omega^2 \left(\omega_{pe}^2 + \omega_{pi}^2 - \omega^2\right) + k^2\vte^2\left(\omega^2- \omega_{pi}^2\right)\right]
\]
\be
\times
\left(\omega_{pe}^2 + \omega_{pi}^2 + c^2 k^2- \omega^2\right)
+\frac{k^2 \omega_{pe}^4 \vte^4}{L_n^2}=0. \label{e11}
\ee
The last term in Eq.~(\ref{e11}) describes coupling of electromagnetic transverse wave from one side, and IA and Langmuir waves from the other. Without this term  the transverse and longitudinal waves propagate without interaction.
 In the ion acoustic frequency range  $\omega^2\ll \omega_{pe}^2, k^2\vte^2$, Eq.~(\ref{e11})  yields the modified
ion acoustic mode
\be
\omega^2=c_s^2\left(k^2 + \frac{m_i}{m_e}\frac{1}{L_n^2}\frac{1}{1+ k^2 \lambda_e^2}\right)  \frac{1}{1+ k^2 \lambda_{de}^2}.\label{e12}
\ee
Here, $\lambda_e=c/\omega_{pe}$, $\lambda_{de}=\vte/\omega_{pe}$.  It is seen that in the limit of small $k$,
the ion-acoustic wave coupled  with the transverse wave in  the presence of density gradient
 has some  cut-off frequency  $\omega_c$, that  goes toward
\be
 \omega_c=\frac{\vte}{L_n}.\label{cut}
 \ee
    \begin{figure}[!htb]
   \centering
  \includegraphics[height=6cm,bb=16 14 260 211,clip=]{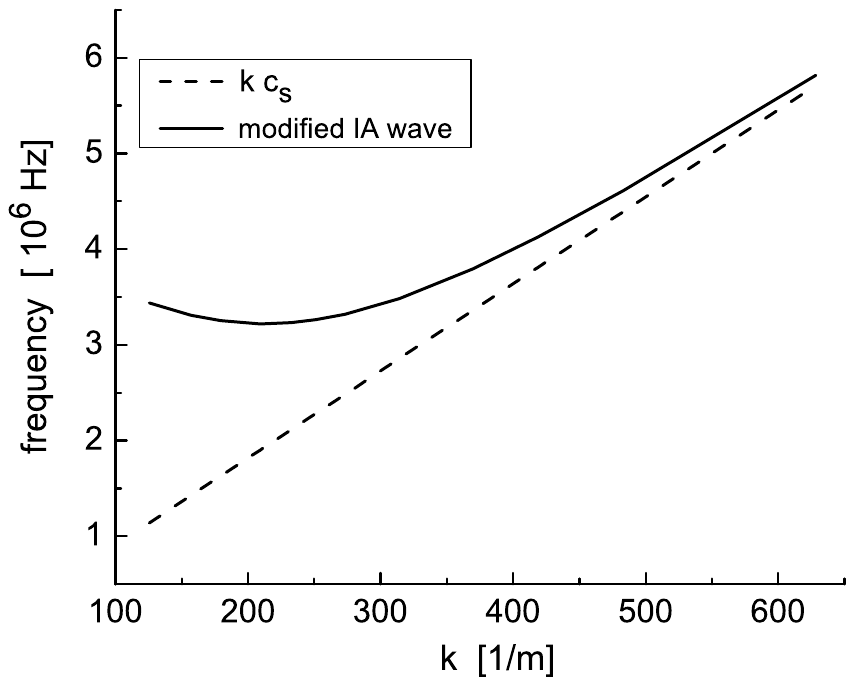}
      \caption{Numerical solution of  Eq.~(\ref{e11}) for the ion acoustic wave in laboratory plasma with backward features and cut-off, both caused by interaction        with transverse light wave in the presence of density gradient.   \label{f1}}
       \end{figure}
In addition, it turns out that the wave group speed may become negative, i.e., the wave becomes backward and  the dispersion curve for the ion acoustic wave may have a trough  in the profile.  A simple condition describing such a situation   can be derived  by neglecting $k^2 \lambda_{de}^2$ with respect to unity.  This yields $\partial \omega/\partial k<0$ provided that
\be
L_n^2<\frac{m_i}{m_e} \frac{\lambda_e^2}{\left(1+ k^2 \lambda_e^2\right)^2}.\label{bac}
\ee
For realistic parameters in  laboratory plasmas this condition may easily be satisfied. In space plasmas this is far less likely because it implies small  values for the density scale length $L_n$ that are certainly possible but are not expected to last long enough so that a wide-spread excitation  of ion acoustic waves  takes place within the life time of such density gradients. However, the cut-off (\ref{cut}) remains as a feature for any plasma.

To  demonstrate possibility for a backward mode, as an example, Eq.~(\ref{e11}) is solved numerically for the laboratory plasma taking cold ions and $L_n=0.1$ m,
$T_e=10^4$ K, $n_0=10^{18}$ m$^{-3}$. The result for the ion acoustic mode is presented in Fig.~\ref{f1}.
It is seen that the mode has backward features for small $k$ and it has a cut-off, all this in agreement with Eqs. (\ref{e12})-(\ref{bac}).

 Here and in Secs. \ref{noniso}, \ref{1c}, \ref{1b} we shall present IA wave only, because properties  of this mode are changed due to coupling. However, it should be kept in mind that the same mechanism of excitation applies to the Langmuir wave as well, though the features of this mode remain almost the same as in the case without coupling. But the energy of both excited waves (IA and Langmuir) is eventually transformed into heat by the same Landau damping effect.

\subsubsection{Perturbations in plasmas with hot ions}\label{noniso}

In the ion-acoustic wave frequency range, the electrons may be taken as isothermal  because $\vte \gg \omega/k=v_{ph}$ (see more in Stix \cite{stix}, and in Bellan \cite{bel}). This considerably simplifies equations without loosing any important physical effect.
 Thus, we set $\delta_{e 1, 2, 3}=0$ and repeat the procedure as above for hot ions, and for their temperature variation described by Eq.~(\ref{en}). Using Eq.~(\ref{e7}) with $\delta_{i 1, 2, 3}\neq 0$ this yields the following wave equation
 \[
 c^2k^2 \vec E_1- c^2 \vec k \left(\vec k \cdot \vec E_1\right) - \omega^2 \vec E_1 + \left(\omega_{pi}^2 + \omega_{pe}^2\right) \vec E_1
 \]
 \[
 - \left\{\!\omega_{pi}^2\vti^2 \left[\! \frac{1+ \delta_{i2}}{\omega_i^2} + \delta_{i3} \left(\! 1+ \frac{k^2\vti^2}{\omega_i^2} (1+ \delta_{i2}) \!\right)\!\right]   + \frac{\omega_{pe}^2 \vte^2}{\omega_e^2}\!\right\}\nabla \left(\nabla\!\cdot \! \vec E_1\right)
 \]
 \[
 -\left\{\!\omega_{pi}^2\vti^2 \left[\! \frac{1+ \delta_{i2}}{\omega_i^2} + \delta_{i3} \left(\! 1+ \frac{k^2\vti^2}{\omega_i^2} (1+ \delta_{i2}) \!\right)\!\right]   + \frac{\omega_{pe}^2 \vte^2}{\omega_e^2}\!\right\}\! \frac{\nabla n_0}{n_0} \nabla\!\cdot\! \vec E_1
 \]
 \be
 - \left[\frac{\omega_{pi}^2 \vti^2}{\omega_i^2} \left(1+ \delta_{i3} k^2 \vti^2\right) + \frac{\omega_{pe}^2 \vte^2}{\omega_e^2}\right]\frac{n_0'}{n_0}\nabla E_{1x}=0. \label{e11a}
 \ee
 The $y$-component again yields one electromagnetic wave, and from the $x, z$-components we obtain the following dispersion equation that  describes
 longitudinal ion-acoustic and  Langmuir waves coupled to another electromagnetic transverse wave with the electric field in $x$-direction:
 \[
\left[\omega^2 - \omega_{pe}^2 - \omega_{pi}^2  - k^2 \left(\frac{5 \omega_{pi}^2 \vti^2/3}{\omega^2- 5 \omega_{pi}^2 \vti^2/3}   +
\frac{\omega_{pe}^2 \vte^2}{ \omega^2- k^2\vte^2}\right)\right]
\]
\[
\times
\left( \omega^2 - \omega_{pe}^2 - \omega_{pi}^2 - k^2 c^2\right) + \frac{k^2}{L_n^2} \left(\!\frac{\omega_{pi}^2 \vti^2}{\omega^2- 5 \omega_{pi}^2 \vti^2/3}   +
\frac{\omega_{pe}^2 \vte^2}{ \omega^2- k^2\vte^2}\!\right)
\]
\be
\times \left(\frac{5 \omega_{pi}^2 \vti^2/3}{\omega^2- 5 \omega_{pi}^2 \vti^2/3}   +
\frac{\omega_{pe}^2 \vte^2}{ \omega^2- k^2\vte^2}\right) =0.
\label{e13}
\ee
In the cold ion limit this equation reduces to Eq.~(\ref{e11}).
Eq.~(\ref{e13}) can be solved numerically for coronal plasma, and for this purpose we take parameters $T=10^6$ K, $n=10^{15}$ m$^{-3}$ and  $L_n=10^5$ m. Note that for these parameters and assuming  the magnetic field $10^{-2}$ T we have $\omega_{pi}/\Omega_i=43$, so indeed the frequency of the transverse wave is far above the gyro-frequency, and the perpendicular dynamics of particles fits into the  unmagnetized plasma model  described previously in the text.

\begin{figure}[!htb]
   \centering
  \includegraphics[height=6cm,bb=17 16 274 217,clip=]{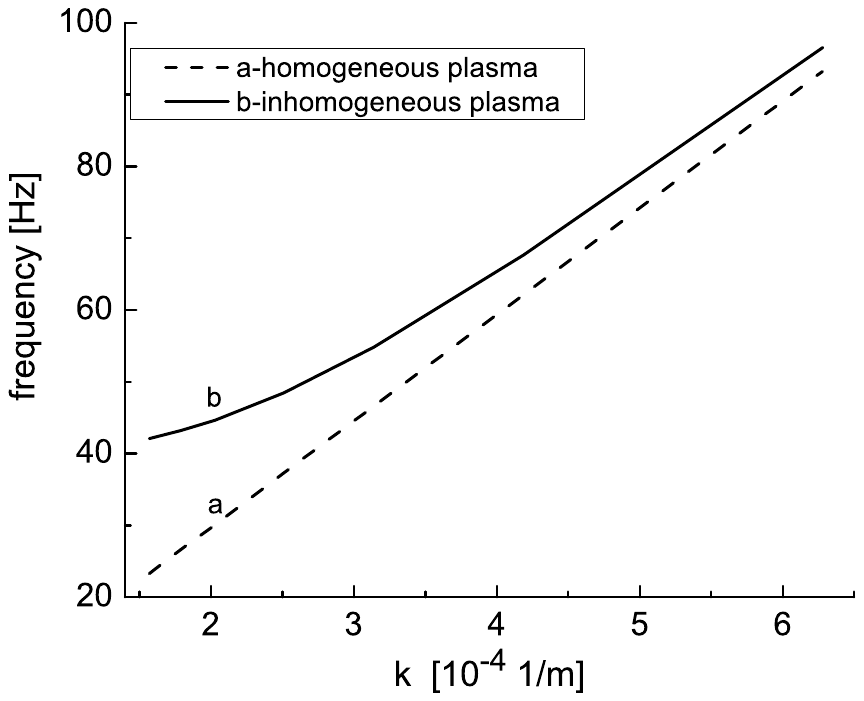}
      \caption{Numerical solution of Eq.~(\ref{e13}) for ion acoustic wave in solar atmosphere. Line a: the usual ion acoustic wave in plasma with hot coronal ions.
      Line b:  the ion acoustic wave  propagating  in the presence of  the transverse light wave in plasma with density gradient.   \label{f2}}
       \end{figure}

 The frequency of the ion-acoustic wave excited   in the presence of the light wave and the  density gradient is presented in Fig.~\ref{f2} (full line), together with the straight (dashed) line  which describes the frequency of  the usual ion-acoustic wave in homogeneous plasma that goes to zero for $k\rightarrow 0$. The cut-off defined by Eq.~(\ref{cut}) in the present case yields the frequency around  $\simeq 40$ Hz. Observe that this is orders of magnitude greater than the cut-off induced  by gravity (Vranjes \cite{v11}) which is around $10^{-3}$ Hz.

\subsection{ Ion Landau damping versus excitation  in the presence of  transverse wave  }\label{1c}

Longitudinal electrostatic ion acoustic perturbations are known to be subject to the Landau damping within which the energy of wave motion is dissipated and transformed into thermal  energy of particles. This  opens up a possibility for precipitation of energy from light waves into heating. This mechanism is expected to work best in magnetic structures where radial density gradients appear as natural features simply due to balance of forces (while the background magnetic field plays no further role for the wave excitation), so the presented new IA wave excitation may be ubiquitous there. The Landau damping is  the  most important ion thermal effect and it  is a purely kinetic phenomenon. However,  there exists  a way to include it even within  the fluid theory using the method suggested several decades ago by D'Angelo et al. (\cite{da79}).   The model implies that in the linearized ion momentum equation a dissipative term    of the shape
 \be
 m_i n_0\mu_i \frac{\partial^2 \vec v_{i 1}}{\partial z^2} \label{ld1}
 \ee
 is added in such a way as to describe well known experimentally verified features of the kinetic Landau damping  (Sessler  and   Pearson \cite{s}, Andersen et al. \cite{a}).
The reason for such a choice, and the properties of the coefficient $\mu_i$ are described in Appendix \ref{apa}. It is interesting to point out that with  an appropriate choice of $\mu_i$ the ion kinetic effects can in fact be described  more accurately than by using the kinetic theory in standard approximate way, i.e.,  by expanding the plasma dispersion function; see more details on this issue in Appendix \ref{apa}.

To describe such a  Landau damping model in the most
transparent way,   we shall assume isothermal but hot ions, and their momentum equation in this case yields
\[
\vec v_{i1}\!=\!\frac{i e  \vec E_1}{m_j \omega_{\mu}} -\frac{i e \vti^2}{m_i \omega_{\mu}\omega_i^2} \nabla \left(\nabla\cdot\vec E_1\right)
-\frac{i e \vti^2}{m_i \omega_{\mu}\omega_i^2} \frac{\nabla n_0}{n_0} \nabla\cdot\vec E_1
\]
\[
-
\frac{i e \vti^2}{m_i \omega_{\mu}\omega_i^2} \frac{n_0'}{n_0} \nabla E_{1x},\quad \omega_{\mu}=\omega + i \mu_i k^2, \quad \omega_i^2=\omega \omega_{\mu} - k^2 \vti^2.
\]
The isothermal electron speed is obtained from Eq.~(\ref{e7}) by setting $\delta_{e1,2,3}=0$.
The derivations are repeated in the same manner as in Sec.~\ref{cold}, and this yields the following  dispersion equation for the $x, z$ components of the electric field
\[
\left(\omega^2 - \omega_{pe}^2 - \omega_{pi}^2\frac{\omega}{\omega_{\mu}}  - \alpha k^2 \right)\! \left(\omega^2 - \omega_{pe}^2 - \omega_{pi}^2 \frac{\omega}{\omega_{\mu}}   - c^2 k^2\right)
\]
\be
+ \frac{k^2 \alpha^2}{L_n^2}\! =\! 0, \label{ld2}
\ee
\[
\alpha= \frac{\omega}{\omega_{\mu}} \frac{  \omega_{pi}^2 \vti^2}{\omega \omega_{\mu} - k^2\vti^2} + \frac{\omega_{pe}^2 \vte^2}{\omega^2- k^2 \vte^2}.
\]
 Dispersion equation (\ref{ld2}) contains  all relevant ion thermal effects, including the Landau damping through the term $\mu_i$ in $\omega_{\mu}$,  and it will  be solved numerically using the expressions for $\mu_i$ from Appendix \ref{ap2}.
    \begin{figure}[!htb]
   \centering
  \includegraphics[height=6cm,bb=17 16 260 217,clip=]{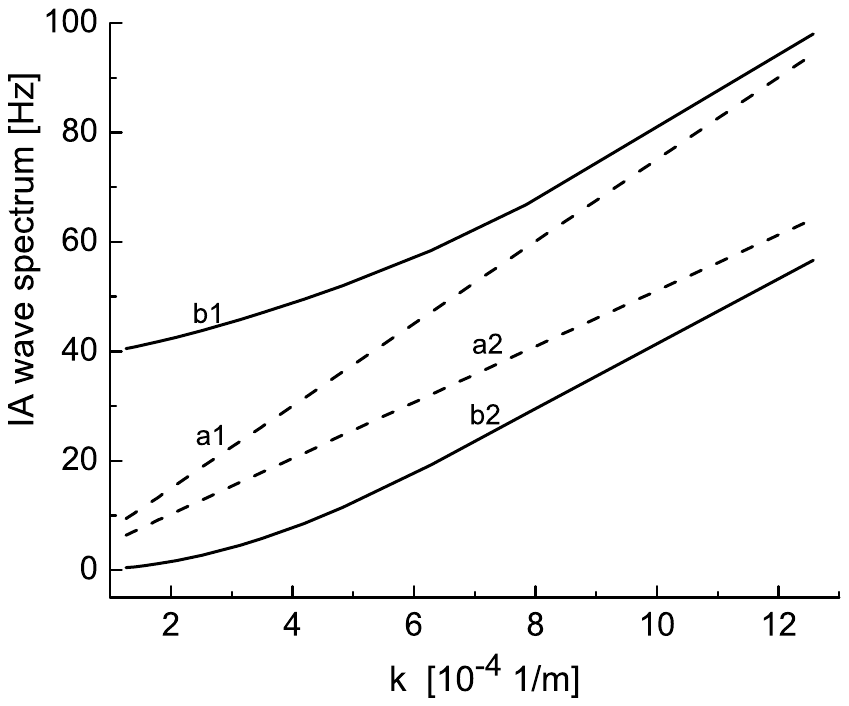}
      \caption{Numerical solution of Eq.~(\ref{ld2}) for the ion acoustic wave spectrum. Dashed lines: the frequency (line a1) and absolute value of Landau damping (line a2) of the ion acoustic wave for homogeneous plasma  with hot coronal ions. Full lines: the   frequency (line b1) and  damping  (line b2) of the ion acoustic wave excited in the presence of   the transverse light wave in inhomogeneous plasma.   \label{f22}}
       \end{figure}
 The result is presented in  Fig.~\ref{f22} for the same  parameters as in Fig.~\ref{f2}. Lines b1 and b2 give, respectively, the frequency and the Landau damping for the  ion acoustic wave   that propagates in the presence of  light waves in plasma with density gradient.  For comparison, the dashed lines a1, a2 give the corresponding frequency and Landau damping for homogeneous  plasma with the same parameters. Observe that the Landau damping given by the line b2 in inhomogeneous plasma is  reduced and it is much smaller than the frequency, in particular for small values of $k$, contrary to the homogeneous plasma where the frequency and damping  become of the same order for small $k$ as it is seen from lines a1, a2. The explanation for this is as follows.

 The Landau effect is most effective in cases when the wave phase speed $\omega/k$ is close to the  thermal speed of the bulk ions. However, from Figs.~\ref{f1}, \ref{f2}, \ref{f22} it is seen that for the ion acoustic wave excited  in the presence of  the light wave,   the frequency for small $k$ is much higher (because of the cut-off effect) than the frequency of the normal ion acoustic wave $k c_s$ in homogeneous plasmas. This implies that in the given range of short $k$ the phase speed is in fact shifted far from the thermal speed, i.e., toward the tail of the ion distribution. Consequently, a considerably smaller amount of ions will satisfy the resonant condition, and the Landau damping must be strongly reduced. These fine features  are seen in Fig.~\ref{f22}; for example, at $k\simeq 1.6$ (in given units) from lines b1, b2 we have  $\omega\simeq 41$ Hz, $\gamma=-0.84$ Hz, so $|\gamma/\omega|=0.02$, while from lines a1, a2 we have $\omega\simeq 12$, $\gamma\simeq 8$ so that $|\gamma/\omega|=0.67$.
The ratio is more than 33 times smaller in the first case!

It may be concluded that the light waves i) not only  help the excitation of  the ion acoustic waves in inhomogeneous environment, but also ii) drastically reduce the Landau damping of these excited IA waves.

\subsection{ Plasma with both temperature and density gradients }\label{1b}

Derivations from Sec.~\ref{1} can be repeated for plasmas which are kept in equilibrium by oppositely oriented density and temperature gradients (Vranjes et al. \cite{vs})
 \be
\frac{\nabla T_0}{T_0}= -\frac{\nabla n_0}{n_0}. \label{t1}
\ee
The equations become very lengthy for hot ions and non-isothermal perturbations, and yet  no new physical effects  appear in comparison
with much more simple isothermal derivations with cold ions. The latter yield the following wave  equation
\[
c^2 k^2 \vec E_1 - c^2 \vec k \left(\vec k\cdot \vec E_1\right)- \omega^2 \vec E_1 + \left(\omega_{pe}^2 + \omega_{pi}^2\right) \vec E_1
-
\frac{\omega_{pe}^2}{L_n} \frac{\vte^2}{\omega_e^2} \nabla E_{1x}
\]
\be
+\frac{\omega_{pe}^2k^2 \vte^4}{\omega_e^4} \frac{\vec e_x}{L_n} \nabla\cdot \vec E_1 - \frac{\omega_{pe}^2\vte^2}{\omega_e^2} \nabla\left(\nabla\cdot \vec E_1\right)=0,  \label{t2}
\ee
\[
 \omega_e^2\!=\!\omega^2\!-\!k^2\vte^2.
 \]
This yields the following dispersion equation for coupled longitudinal and light waves:
\[
\left(\omega^2- k^2 \vte^2\right)^2 \left[ \omega^2 \left(\omega_{pe}^2 + \omega_{pi}^2 - \omega^2\right) + k^2\vte^2\left( \omega^2- \omega_{pi}^2 \right)\right]
\]
\be
\times \left(\omega_{pe}^2 + \omega_{pi}^2 + c^2 k^2- \omega^2\right)
=\frac{k^4 \omega_{pe}^4 \vte^6}{L_n^2}. \label{t3}
\ee
Eq.~(\ref{t3})  is slightly different than Eq.~(\ref{e11}), however  in the ion acoustic frequency range it  yields the same dispersion equation (\ref{e12}) as before.

Consequently, the ion acoustic wave excitation in the presence of   the light wave remains similar even for a rather different equilibrium  (\ref{t1}). The physics of coupling will not change by including  hot ion effects. The  presented mechanism of excitation  can thus act in various geometries, with or without the external magnetic field.

\subsection{Refraction and focusing of transverse wave by background  density gradient:  accumulation of wave  energy in regions with lower density}\label{foc}

In a less ideal picture,  the oscillating electromagnetic field may in fact have electric field at an arbitrary angle with respect to the density gradient. Nevertheless, the mechanism described in Fig.~\ref{mech}   will work as long as this electric field is not strictly orthogonal to $\nabla n_0$. This implies that in real situations the described coupling between transverse and longitudinal waves may be widespread  because electromagnetic waves are indeed omnipresent, and they propagate in any direction independent on the magnetic field geometry.

On the other hand, the phase speed of the transverse wave is greater than the speed of light
\be
v_{ph}=\left(c^2 + \frac{\omega_p^2}{k^2}\right)^{1/2},
\label{fo1}
\ee
which implies that the coupling with IA perturbations may take place practically   instantaneously along large sections of a magnetic flux tube through which the EM  wave propagates.

\begin{figure}[!htb]
   \centering
  \includegraphics[height=6cm,bb=15 14 235 210,clip=]{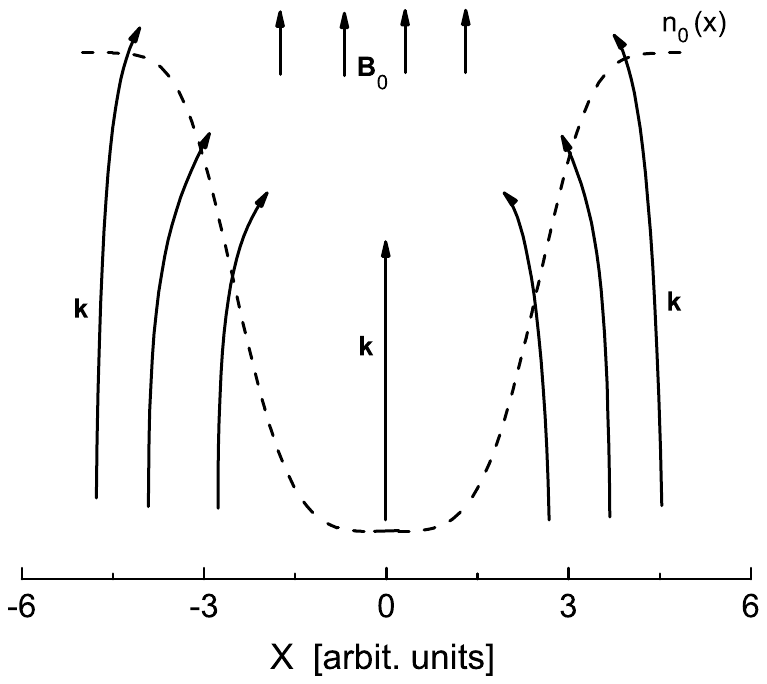}
      \caption{Refraction and focusing of transverse electromagnetic wave  with the wave number $\vec k$  in plasma with density gradient.    \label{prof}}
       \end{figure}

From the given phase speed, it is also seen that the wave with certain  frequency propagates faster in a more dense environment. In a magnetic flux tube, due to pressure balance, one may expect reduced density inside the tube and enhanced at the edge. The refraction index of the transverse wave is $N=c/v_{ph}<1$, so that wave front initially propagating along such a magnetic flux tube will bend, and there will be refraction and focusing of the wave towards the center of the tube cylinder. This also implies that the direction of the wave electric field will continuously change its angle with respect to the magnetic field vector, but the excitation of the IA wave will still be taking place as explained earlier.   This refraction and focusing of the EM wave is depicted in Fig.~\ref{prof}, where the wave initially propagates along the magnetic field and it has  electric field vector in the direction of the background density gradient. The wave front becomes distorted so  that the wave is directed towards the region of lower plasma density.

The consequences of this effect may be  numerous.  In laboratory situations this effect is used for plasma heating by CO$_2$ lasers (Chen \cite{ch}); an elongated plasma with lower density on the axis acts as a lens, the radiation converges  towards the axis and becomes trapped inside the plasma.
The same lens effect is behind radio communications around the globe; in this case waves are reflected from ionospheric layers and are therefore able to travel around the globe.

A similar physical mechanism is behind well known self-focusing of strong electromagnetic beams.  In this case the nonlinear ponderomotive force pushes plasma out of the beam and the phase speed inside the beam is smaller as compared to outer region so that the beam becomes focused.

The principles of linear fusion reactor (i.e., plasma heating) described above, which are known and used in rather unfavorable conditions in typical laboratory plasmas,  may naturally  be in action in very elongated magnetic flux tubes in the solar corona where conditions for such a heating are much more favorable. Hence, such a focused electromagnetic transverse wave will more effectively  support excitation of   the ion acoustic waves by the mechanism described previously in the text, and  it can also  heat inner regions of magnetic flux tubes through  nonlinear wave coupling  mentioned in Sec.~1.

Such mechanisms, although well explored and used in practice in laboratory plasmas, have not been studied in the problem of coronal heating. Focused EM or light waves imply larger wave amplitudes inside a region of less dense plasma (e.g., in a flux tube), and in such circumstances   nonlinear effects may become important, like the parametric decay instability mentioned in Sec.~\ref{s1}. In this case the EM wave energy is converted into plasma wave energy and the later is then transferred directly to electrons through Landau damping. This mechanism is standardly used in laser fusion experiments, e.g, in the iodine laser heating. The required EM wave amplitude for this phenomenon is known (Bellan \cite{bel}), and it reads
\be
E_{{\sss EM}}=4 \frac{(\omega_1 \omega_2 \gamma_1\gamma_2)^{1/2}}{\Lambda}.
\label{fo2}
\ee
Here,  $ \Lambda$ is the nonlinear coupling coefficient
\[
\Lambda=\frac{e k_1 \omega_{pe}}{(m_i m_e \omega_2 \omega_3)^{1/2}}, \quad \omega_1=\omega_3-\omega_2, \quad \vec k_1=\vec k_3-\vec k_2,
\]
indices $1, 2, 3$ refer to IA, PL, and EM waves, respectively.  $\gamma_1, \gamma_2$ are linear damping rates for IA and PL modes, which in application to coronal plasmas  are just the usual Landau damping rates.

Very similar is nonlinear excitation of IA wave through so called parametric backscattering instability which involves an  EM pump wave that excites one IA and another EM wave moving in opposite direction. This requires EM pump wave amplitude same as above (\ref{fo2}),  but with slightly different coupling constant
\[
\Lambda=\frac{e k_1 \omega_{pe}}{\omega_3(m_i m_e)^{1/2}}.
\]
For this case the IA damping rate  is the same as above, while for the EM wave (which suffers no Landau damping) the damping can only be due to electron-ion collisions (typically on the order of 1 Hz, Vranjes and Poedts \cite{v06}) and it reads (Chen \cite{ch}) $\gamma_2=\nu_{ei} \omega_{pe}^2/(2 \omega_2^2)$. So this nonlinear coupling is accompanied with possible energy channeling into IA waves and further with ion heating due to Landau effect.

This shows that the IA wave is excited in both linear and nonlinear regimes.   In the linear regime the EM wave introduces a triggering mechanism only by producing perturbed longitudinal electric field due to the density gradient. In other words it acts as a catalyst,  yet  it does not actually drive the IA wave.  In both cases it is due to the density gradient.   In the linear regime this is in the region of density gradient (e.g., mainly in the outer region of a magnetic tube), while in the nonlinear regime this is the central region of the tube to which the EM wave is focused due to the same density gradient.

\section{Growing ion acoustic wave in solar corona}\label{gro}

\subsection{Instability in permeating plasmas}\label{kin}

 Details of the mechanism b) introduced in Sec.~\ref{s1}  are described in  Vranjes et al. (\cite{perm}), and in  Vranjes (\cite{v9}). This is a  kinetic derivation  within which the plasma distribution function for the species $j$ is used in the form
\be
f_{j0}=\frac{n_{j0}}{(2\pi)^{3/2}\vtj^3} \exp\left\{-\frac{1}{2
\vtj^2}\left[v_x^2 + v_y^2 + (v_z- v_{j0})^2\right]\right\},
\label{k4} \ee
where $n_{j0}=const$,   $v_{j0}$ is some equilibrium speed.  This is used in the linearized
Boltzmann kinetic equation considering two plasmas, one static   and another one  inflowing from some other region,  denoted with superscripts $s$ and $f$, respectively. The quasi-neutrality conditions in the equilibrium are naturally satisfied in both plasmas  separately:
\be
n_{fi0}=n_{fe0}=n_{f0}, \quad n_{si0}=n_{se0}=n_{s0}, \label{k5}
 \ee
and we shall study  longitudinal electrostatic perturbations of the shape $\sim \exp(-i
\omega + i k z)$. Here, the direction $z$ is arbitrary and in principle it should not be identified with  the magnetic field direction as in the previous  sections.  Using the Amp\`{e}re law
after a few  straightforward steps we obtain the dispersion equation
for  the ion acoustic wave  (Vranjes et al. \cite{perm},  Vranjes \cite{v9}):
\[
 1+ \frac{1}{k^2 \lambda_d^2} -
\frac{\omega_{psi}^2}{\omega^2} - \frac{3 k^2 \vtsi^2
\omega_{psi}^2}{\omega^4} + i \left(\frac{\pi}{2}\right)^{1/2} \left[\frac{\omega
\omega_{pse}^2}{k^3 \vtse^3}\right.
\]
\be
\left.
 + (\omega- k
v_{f0})\!\left(\frac{\omega_{pfe}^2}{k^3 \vtfe^3} +
\frac{\omega_{pfi}^2}{k^3 \vtfi^3}\!\right)
+
\frac{\omega\omega_{psi}^2 }{k^3 \vtsi^3}\! \exp\left(\!-
\frac{\omega^2}{2 k^2 \vtsi^2}\right)\!\right]\!=\!0. \label{k8} \ee
In the derivation of  Eq.~(\ref{k8}) the following  limits  are used
\be
k \vtsi \ll |\omega|\ll k \vtse, \quad |\omega - k v_{f0}|\ll k
\vtfe, \,\, k\vtfi, \label{c2} \ee
The notation is as follows: $1/\lambda_d^2=1/\lambda_{dse}^2 + 1/\lambda_{dfe}^2 +
1/\lambda_{dfi}^2$,  $\lambda_{dse}=\vtse/\omega_{pse}$ etc., and  $ \vtsj^2=\kappa
T_{sj}/m_j$, $ \vtfj^2=\kappa T_{fj}/m_j$.

To demonstrate the instability (growing) of the ion acoustic wave we take the following set of parameters:  the temperature of all species is set to $10^6$ K, and the number density  for all species is taken as $n_0=10^{13}$ m$^{-3}$. Such a choice is taken only to formally and easily satisfy the conditions (\ref{c2}) used for the expansions performed above which  have yielded the dispersion equation (\ref{k8}). Clearly, any other combinations of the temperatures and the number densities are possible, however in order to remain self-consistent this would imply solving the general dispersion equation  [given in Vranjes et al. \cite{perm},  Vranjes \cite{v9})  instead of Eq.~(\ref{k8}]. This general one  contains complex functions of complex arguments and such a task could only be done numerically.

If in addition we take a relatively  weak magnetic field $B_0=5\cdot 10^{-4}$ T, for the assumed parameters we have the gyro-frequency and the gyro-radius for protons $\Omega_i\simeq 48$ kHz, $\rho_i\simeq 1.9$ m. Hence, we may search for the ion acoustic mode with frequency above $\Omega_i$, which will therefore be able to propagate at almost any angle with respect to the magnetic field, and the magnetic field will not affect the dynamics of ions. Note that this is not essential for the IA wave instability which shall be demonstrated below, but it may be of importance  for associated stochastic heating (see below in Sec.~\ref{stoc})  towards  which this instability will eventually lead.

\begin{figure}[!htb]
   \centering
  \includegraphics[height=6cm,bb=20 17 278 217,clip=]{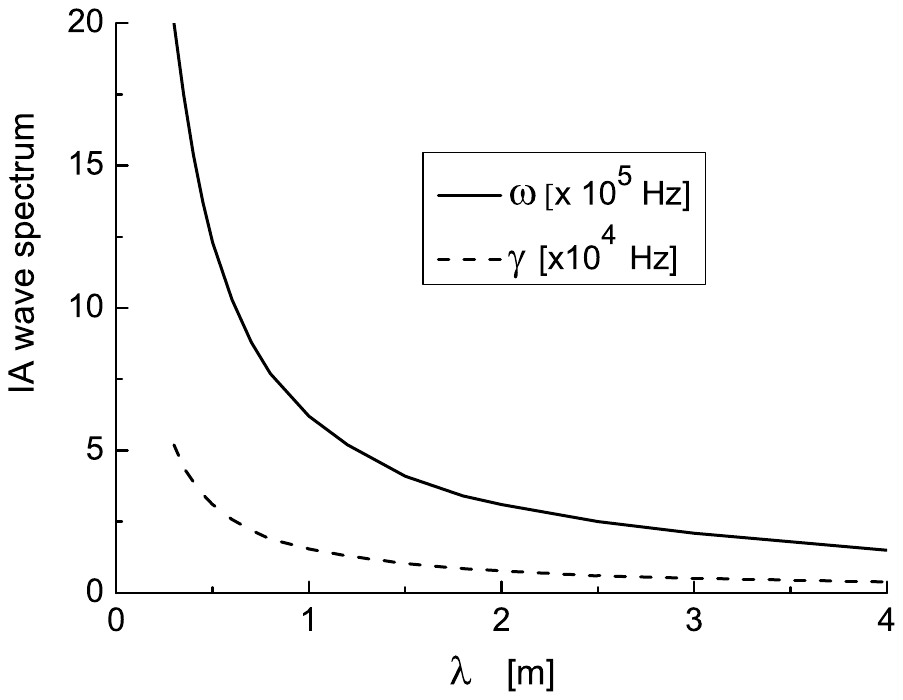}
      \caption{Growing ion acoustic wave in permeating coronal plasmas with the speed of the flowing quasi-neutral plasma  with respect to the static plasma $v_{f0}=1.9 c_s$.   \label{fk1}}
       \end{figure}

For waves propagating parallel to the magnetic field vector the particles will not feel its effect anyhow. For this case  dispersion equation (\ref{k8}) is solved first in terms of the wavelength and the result is presented in Fig.~\ref{fk1} for the speed of the flowing plasma $v_{f0}=1.9 c_s$. The figure shows a growing  (unstable) mode. This is a remarkable result because the ion acoustic mode is believed to be mostly impossible to excite in plasmas with hot ions. In the present case the source of the instability is the energy stored in the flowing plasma.

Equally remarkable feature is a rather low instability threshold
which is more than one order of magnitude lower than  the threshold for electron current driven IA wave instability mentioned in Sec.~\ref{s1}.  Observe also that this instability in permeating plasmas is essentially current-less, so there is no problem of consistency related to assumed homogeneity  of magnetic field as in the case of current-carrying plasma.

\begin{figure}[!htb]
   \centering
  \includegraphics[height=6cm,bb=17 17 273 210,clip=]{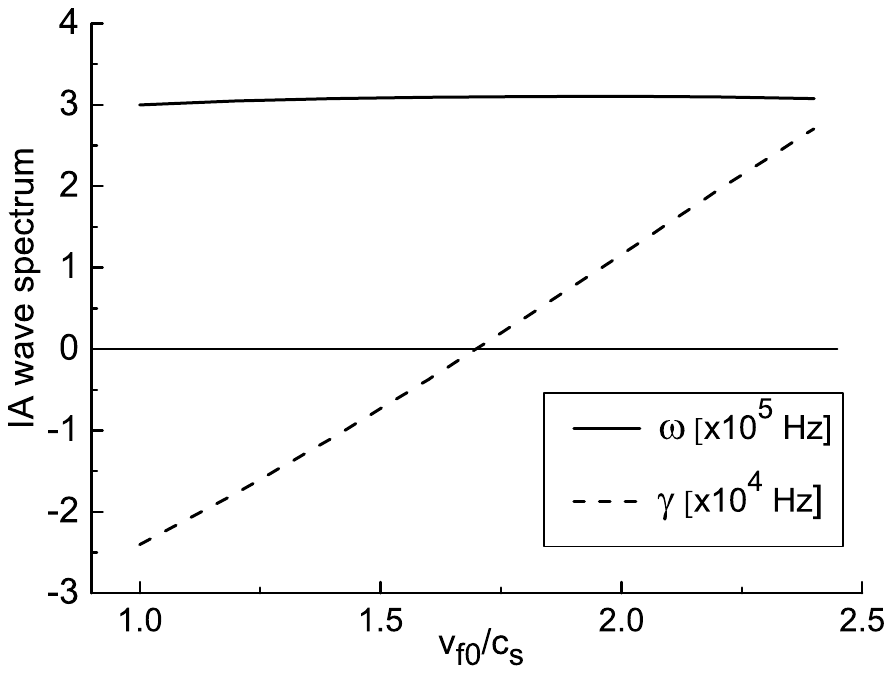}
      \caption{Transition from damped to growing ion acoustic mode in permeating solar plasmas  for $\lambda=2$ m.\label{fk2}}
       \end{figure}

The transition from the usual Landau damping to growth is demonstrated in Fig.~\ref{fk2} where the wavelength is fixed to $\lambda=2$ m (c.f. Fig.~\ref{fk1}) and the speed of the flowing plasma is allowed to change. The mode becomes growing at around $v_{f0}/c_s=1.7$. For the parameters used here this implies the flowing plasma speed of around 150 km/s, which is for example far below the solar wind speed. Hence, the ion acoustic wave can become unstable (growing) even in an environment with hot ions like the solar corona  where it is typically expected to be strongly damped.

For an oblique propagation at an angle $\theta$ with respect to the magnetic field, following derivations in Alexandrov et al. (\cite{aleks}) it turns out that the dispersion equation does not change much. For un-magnetized ions this is self-evident and for electrons the explanation is as follows.
The longitudinal electron permitivity includes the summation
  \[
  \sum_{n=-\infty}^{+\infty} \frac{\omega}{\omega- n \Omega_e} A_n(k_{\bot}^2 \rho_e^2) Z(b_e)
  \]
  where $b_e=(\omega- n \Omega_e)/(k_z \vte)$,  $A_n(\eta)\equiv e^{-\eta} I_n(\eta)$,  $I_n(\eta)$ are the Bessel functions of the second kind, and $\omega$ may include the Doppler shift as above. Observe $k_z$ in these expressions instead of $k$ as compared with  Eq.~(\ref{k8}). In the limit $|(\omega- n \Omega_e)/(k_z \vte)|\ll 1$ the appropriate expansion for $Z(b_e)$ is used,  and as a result the term containing the frequencies vanishes, while  in the same time the summation becomes $\sum_n A_n(\eta)=1$.
  The terms driving the instability in Eq.~(\ref{k8})  are of the form $1- k v_{f0}/v_{ph}$ where $v_{ph}=\omega/k$,  while now for the oblique propagation $k$ becomes $k_z$ and the whole term  turns out to be  replaced with
\be
\frac{1}{\cos \theta}\left(1- \frac{v_{f0}}{v_{ph}} \cos \theta\right). \label{ang}
\ee
Thus, for not so large angle of propagation  the instability features of the mode presented in Figs.~\ref{fk1}, \ref{fk2} will not substantially change in the case of  oblique propagation, while the heating becomes more effective (Smith and Kaufman \cite{sk1}, \cite{sk2}).
Taking the angle of propagation of 30 degrees, in view of (\ref{ang}) the growth rate will be increased by a factor $1.16$.

\subsection{On stochastic heating by the ion acoustic wave}\label{stoc}

The theory presented in Sec.~\ref{kin} is suitable for stochastic heating whose features are as follows. In the case of a particle moving in the field of a sinusoidal wave which propagates at an angle with respect to the magnetic field vector
$\vec B_0=B_0\vec e_z$, the resonance appears  if
\be
\omega- k_z v_z=l \Omega, \quad l=0, \pm 1, \pm 2\cdots,   \label{s2}
\ee
where, $v_z$ is the particle speed in the direction of the magnetic field.     Represented in the phase space $v_z, z$ the resonances have a narrow  width
if the wave amplitude is  infinitesimal and there can be no overlapping in that case. For larger wave amplitudes overlapping of resonances may take place, therefore the  particle may satisfy two resonant conditions in the same time and its motion becomes random (stochastic), with its energy increased in the process.

In terms of the wave amplitude, for the high frequency  IA wave ($\omega>\Omega_i$) the  condition for stochastic heating  reads
 \be
 e k_z^2 \phi_1 |J_l(k_{\bot} \rho_i)|/m_i\geq \frac{\Omega_i^2}{16}, \quad l=0, \pm 1, \pm 2\cdots \label{sto}
 \ee
 Here, $k_z, k_{\bot}$ are the parallel and perpendicular wave-number components with respect to the magnetic field vector, $\phi_1$ is the wave potential,
 $\rho_i, \Omega_i$ are the ion gyro-radius and gyro-frequency, $J_l$ is the Bessel function, and $l$ is any integer which describes higher harmonics. The heating act mainly on the ions in the tail of the distribution function.
 Similarly, for the ion cyclotron (IC)  and the lower-hybrid (LH) waves the necessary threshold (Karney and Bears \cite{kar}, Karney \cite{kar2}) can be written as $E/B_0>(\Omega_i/\omega)^{1/3} (\omega/k_{\bot})/4$. These modes  stochastically heat mainly the bulk plasma.
The condition for the onset of the stochastic behavior in the case of the IA wave (\ref{sto}) is obtained simply from the condition that the sum of two half-width of adjacent resonances exceeds the separation between them (Smith and Kaufman \cite{sk1}, \cite{sk2}).  The heating is typically very rapid, with the rates comparable to the gyro-frequency $\Omega_i$, and it is also very efficient for the IA waves that propagate at an angle with respect to the magnetic field vector. Clearly, this implies the wave frequency exceeding the gyro-frequency $\Omega_i$. Following Smith and Kaufman (\cite{sk2}) the optimal case is when the wave frequency is around 3-4 times the gyro-frequency.  This is the reason why in Sec.~\ref{kin} we restricted analysis to  high frequency IA waves; the instability obtained there will in fact remain for a rather wide range of wave-lengths and frequencies. Note also that, as discussed in Smith and Kaufman (\cite{sk2}) this does not exclude the heating in the low frequency regime $\omega<\Omega_i$, although it will not be as  strong as in the previous case.

The heating implies that the wave loses its energy, and this should be compensated by some moderate instability that may keep the wave amplitude above the required level (\ref{sto}). The instability which appears in interpenetrating plasmas, presented in Sec.~\ref{kin}, may clearly serve this purpose.
Following the prescription for the most effective heating given in Smith and Kaufman (\cite{sk2}), the model should be restricted to weakly  collisional plasmas and to short perpendicular wavelengths.
From Eq.~(\ref{sto}) and for parameters assumed in Sec.~\ref{kin} we  have the critical amplitude of the wave  potential of only $\phi_c=0.41$ V. This implies that the ratio $e \phi_c/(\kappa T)=0.005$ is still a rather small quantity, thus  the level of electrostatic fluctuations is rather low, yet strong enough to cause a rapid and substantial heating as predicted by Smith and Kaufman (\cite{sk2}).

\section{Summary}

In this work a  new kind of ion acoustic wave is derived, with rather different features as compared to ordinary IA wave in homogeneous environment.
The mode can   spontaneously be excited    provided  the simultaneous  presence of  the transverse electromagnetic wave and density gradient, it has a cut-off and it can be backward.
The excitation and coupling to the transverse wave is a purely linear effect. The cut-off is   quantitatively and quantitatively  very different from the IA wave cut-off that  follows from the stratification of plasma in external gravity field (Wahlberg  and  Revenchuk \cite{wah}, Vranjes \cite{v11}), and that also implies a  plasma with  inhomogeneous density. This new kind of IA wave in hot ion plasmas is in principle damped due to the Landau effect, however it is shown that the damping is drastically reduced due to the presence of the cut-off which pushes the resonance towards the tail of the distribution function where the amount of resonant ions is much smaller. These effects are presented quantitatively for the coronal plasma parameters. The dissipation still remains, and it in fact opens up a possibility for a direct transfer of radiation energy (from transverse electromagnetic light waves) to plasma internal energy and heating (by excitation of the IA wave and its consequent dissipation).

We stress that for the purpose of using the Landau damping within the fluid theory, we have presented an excellent and very useful method suggested many years ago by D'Angelo et al. (\cite{da79}), but completely overlooked by scientific community.

One new kind of {\em instability} of the ion acoustic wave  which develops in permeating plasmas,  discovered previously in  Vranjes et al. (\cite{perm}),  is also discussed in application to the solar corona.  It is shown that this instability
can be rather effective (it has  much lower threshold than the instability in usual electron-ion plasma with one component moving with respect to the other),  it is current-less, and it can excite the IA wave even in environments with hot ions like the solar corona. We have discussed high-frequency range of the IA waves
which consequently may propagate practically at any angle with respect to the background magnetic field. Such IA waves are known to cause stochastic heating and this effect is discussed in application to solar corona. This  mechanism of stochastic heating is completely different than the dissipative heating due to Landau damping mentioned above.

In the presence of collisions the  Landau damping can be  reduced (Ono and Kulsrud \cite{ono}) but this may be expected only for wave-lengths considerably exceeding the mean free path of particles. So in the case of high frequency IA waves discussed in Sec.~\ref{gro} this is not expected to play any role. The same holds for the wave-lengths used in Sec.~\ref{1}, but in principle for larger wave-lengths the Landau dissipation may require longer time intervals.

 %\vfill

\section*{Appendix: Landau damping of ion acoustic waves in   fluid theory } \label{apa}

\subsection*{Elements of kinetic description}

Within the two-component kinetic theory the Landau damping is well known. Here, we give a few details only in order to compare it with fluid description presented below in Sec. \ref{ap2}.  Small electrostatic perturbations of the form
$\exp(-i \omega t + i k z)$ yield  the  dispersion equation

\be
1+ \sum_j \frac{1}{k^2 \lambda_{dj}^2} [1- Z(b_j)]=0, \label{k6}
 \ee
\[
b_j=\frac{\omega- k v_{j0}}{k \vtj}, \quad \lambda_{dj}=\frac{\vtj}{\omega_{pj}}, \quad \omega_{pj}^2= \frac{q_j^2
n_{j0}}{\varepsilon_0 m_j},
\]
\[
Z(b_j)=\frac{b_j}{(2 \pi)^{1/2}} \int_c \frac{\exp(-
\zeta^2/2)}{b_j- \zeta} d\zeta.
\]
 The dispersion equation (\ref{k6}) describes the
plasma (Langmuir) and the ion acoustic oscillations, and $\zeta=v_z/\vtj$.

 For the present case with electrons and ions, and in the limit
 \be
v_{{\sss T}i} \ll  |\omega/k| \ll  v_{{\sss T}e}, \label{con1} \ee
the standard  expansions may be  used ${\cal Z}  \approx - i
(\pi/2)^{1/2}\omega/(k v_{{\sss T}e})$, and ${\cal Z} \approx 1+ (k
v_{{\sss T}i}/\omega)^2 +
 3 (k v_{{\sss T}i}/\omega)^4
 + 15 (k v_{{\sss T}i}/\omega)^6  + \cdot\cdot\cdot
- i (\pi/2)^{1/2}[\omega/(k v_{{\sss T}i})]\exp\{- [\omega/(k
v_{{\sss T}i})]^2/2\}$, where $\omega=\omega_r + i \gamma$ is
complex and its real part is assumed to be much larger than its
imaginary part. The procedure yields the dispersion equation
\[
1+ \frac{\omega_{pe}^2}{k^2 v_{{\sss T}e}^2} \left[1+ i
(\pi/2)^{1/2} \frac{\omega}{k v_{{\sss T}e}} \right]
 - \frac{\omega_{p i}^2}{k^2 v_{{\sss T}i}^2} \left\{ \frac{k^2
v_{{\sss T}i}^2}{\omega^2} \right.
\]
\be
\left.
 +  \frac{3 k^4 v_{{\sss T}i}^4}{\omega^4}
\right. \left.
  -
i (\pi/2)^{1/2} \frac{\omega}{k v_{{\sss T}i}}
\exp\left[-\omega^2/(2 k^2 v_{{\sss T}i}^2)\right]\right\}
=0.\label{ap3} \ee
We introduce  $\Im \triangle(\omega, k)$ and $\Re \triangle(\omega,
k)$ denoting the imaginary and real parts of Eq.~(\ref{ap3}), respectively.
Setting the real part $\Re \triangle(\omega, k)=0$ yields the
approximate expression for the  ion acoustic (IA) spectrum
 $\omega_r^2\simeq k^2 (c_s^2 + 3 \vti^2)$,
$c_s^2=\kappa T_e/m_i$. The approximate Landau damping of the wave
is obtained from
  \[
\gamma_{app} \simeq - \Im \triangle(\omega, k)/[\partial \Re
\triangle(\omega, k)/\partial \omega]_{\omega
 \simeq  \omega_r}
\]
\be
 =- k c_s (\pi/8)^{1/2} \tau^{3/2} (1+ 3/\tau) \exp[-(\tau + 3)/2].
\label{ap4} \ee
Here, $\tau=T_e/T_i$ and  we have assumed massless electrons and singly
charged ions, i.e., the Landau damping is due to the ions only.
Otherwise, the electron  contribution to the Landau damping
$-\omega_r (\pi/8)^{1/2} (z_i^3 m_e/m_i)^{1/2}$, would appear in addition
to the previously given Landau damping term,  where $z_i$ is the ion
charge number.

    \begin{figure}[!htb]
   \centering
  \includegraphics[height=6cm,bb=20 16 285 217,clip=]{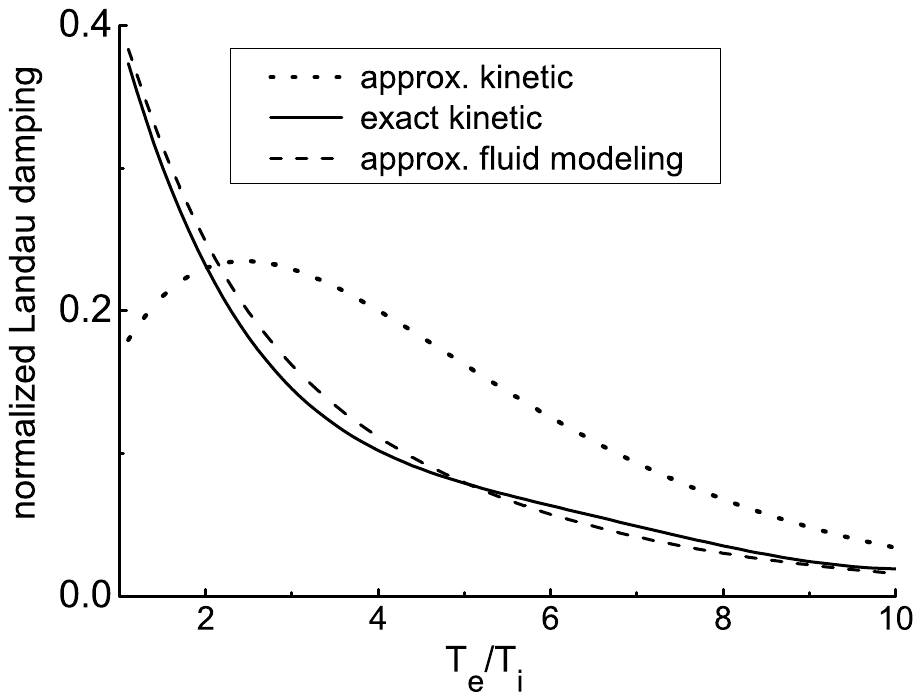}
      \caption{The absolute value of the Landau damping of
the IA wave (normalized to $\omega_r$) in terms of $\tau=T_e/T_i$. \label{land}}
       \end{figure}
In Fig.~\ref{land}, we give the absolute value of the approximate (dotted
line) Landau damping (\ref{e4})  (normalized to $\omega_r$) in terms
of $\tau$, together with the absolute value of {\em the exact}
normalized Landau damping (full line) that may be obtained
numerically from the general dispersion law (\ref{k6}). The nature
of the dashed line will be given in Sec.~\ref{ap2} below. It is seen
 that in the given range of $\tau$ the approximate kinetic
expression (\ref{ap4}) yields inaccurate values for the damping. For
$\tau=1$, which is of interest for the solar corona, the  exact
Landau damping $\gamma_{ex}=0.394 \omega_r$ is larger by about a
factor 2 than the approximate kinetic Landau damping $\gamma_{app}$.

For practical purposes, note that the {\em exact} kinetic Landau damping may
conveniently be expressed in terms of $\tau$ by the following
polynomial [cf.\ Chen (\cite{ch})]:
  \[
|\gamma_{ex}|/\omega_r \simeq  0.681874541  -0.3697636428549\,\tau
\]
\[
 +0.09345888549671\,\tau^2   -0.01203427025053\,\tau^3
\]
\be
+0.0007523967677\,\tau^4-0.000018\,\tau^5.\label{ex}
  \ee
The reason for the difference between $\gamma_{ex}$ and
$\gamma_{app}$ follows from the fact that, in the limit
$\tau\rightarrow 1$, the assumptions used to obtain  the approximate
kinetic damping become violated. One is thus supposed to  use  the
rather inconvenient general dispersion equation (\ref{k6}), which
contains an integral. Below, we show that this problem may be
circumvented by using the two-fluid theory and a modeled
fluid-Landau damping.

\subsection*{Two-fluid description}\label{ap2}

We use the linearized momentum equation for the  general species $\alpha$
in a quasi-neutral plasma
\be
m_\alpha n_0 \frac{\partial v_{\alpha 1}}{\partial t} = - q_\alpha
n_0\frac{\partial \phi_1}{\partial z} - \kappa T_\alpha
\frac{\partial n_1}{\partial z} + \mu_{{\sss L}}
\frac{\partial^2 v_{\alpha 1}}{\partial z^2}.\label{ap5} \ee
The term with $\mu_{{\sss L}}$ is the one introduced for the
first time  by D'Angelo et~al.\ (\cite{da79}) in order to describe the
Landau damping effect on the fast solar wind streams with a
spatially varying ratio $\tau (= T_e/T_i)$ of the electron and ion
temperatures. It is
chosen in such a way to quantitatively describe  the known properties of the Landau
effect. These include the fact that the ratio $d=\delta/\lambda$,
between the attenuation length $\delta$ and the wavelength, is
independent of the wavelength and  the plasma density, and dependent
on $\tau$ in a prescribed way  (Sessler  and   Pearson \cite{s}, Andersen et al. \cite{a}). These
requirements appear to be fulfilled by
\be
\mu_{{\sss L}} =\frac{m_\alpha n_0v_s \lambda}{2 \pi^2 d}.
\label{m} \ee
Here,  $v_s^2=c_s^2 + \vti^2$, while $d(\tau)$  satisfies a curve
which is such that the attenuation is strong at $\tau\approx 1$ and
weak for higher values of $\tau$. It turns out that a sufficiently good
choice for $d(\tau)$ is:
\[
 d(\tau)=0.2750708 + 0.0420737789\,\tau +0.0890326\,\tau^2
\]
\[
 -0.011785\,\tau^3+ 0.0012186\,\tau^4.
 \]
Observe that $d(1)\simeq 0.4$. We also use  the ion continuity and
in the massless electron limit from the electron momentum, we obtain
just the Boltzmann distribution for electrons. This yields the
dispersion equation for the Landau-damped IA wave
\be
 \omega^2 + i \mu_i \omega k^2 - k^2 v_s^2=0, \quad \mu_i= \frac{\mu_{{\sss L}}}{m_i n_0}. \label{a6}
 \ee
For the complex frequency $\omega=\omega_r+ i \gamma_f$  we obtain
 \be
\gamma_f=-\mu_i k^2/2=- v_s/(\lambda d), \quad \omega_r^2= k^2
v_s^2- \mu_i^2 k^4/4. \label{reim}
 \ee
This shows that the mode oscillation frequency $\omega_r$ is reduced as compared
to the ideal plasma case, i.e.\ without Landau damping. It is appropriate
now to compare this fluid-modeled Landau damping $\gamma_f$ with the
exact kinetic damping given earlier. The normalized absolute
fluid-modeled Landau damping $|\gamma_f/\omega_r|\simeq 1/(2 \pi d)$
is presented by the dashed line in Fig.~\ref{land}. The difference between
$\gamma_f$ and $\gamma_{ex}$ at $\tau=1$ is only $0.008$. Clearly,
the presented fluid-modeled Landau damping is in fact much more accurate
than the approximate kinetic expression (\ref{ap4}).

%\begin{acknowledgements}

%ggg

%\end{acknowledgements}


\begin{thebibliography}{}

\bibitem{aleks}  Alexandrov, A. F., Bogdankevich, L. S., \& Rukhadze, A. A.  1984, Principles of Plasma Electrodynamics (Springer-Verlag, Berlin Heidelberg)

\bibitem{alf1} Alfv\'{e}n, H.  1947, MNRAS,
    107, 211

    \bibitem{a} Andersen, H. K.,  D'Angelo, N.,   Jensen, V. O., Michelsen, P.,  $\&$ Nielsen, P. 1968,
 Phys. Fluids, 11, 1177


\bibitem{bel} Bellan, P. M. 2006, Fundamentals of  Plasma Physics (Cambridge Univ. Press, Cambridge)

\bibitem{belo} Bello Gonz\'{a}lez, N. et al.,  2011, ApJL, 723, 134

\bibitem{bir1} Biermann, L.  1948, Zs. f. Ap, 25, 161


\bibitem{ch} Chen, F. F. 1984, Introduction to Plasma Physics and Controlled Fusion (Plenum Press, New York)

\bibitem{da1} D'Angelo, B.  1968, ApJ, 154, 401

\bibitem{da79} D'Angelo, N.,  Joyce, G.,  $\&$  Pesses, M. E. 1979,  ApJ,  229, 1138


\bibitem{dov} Doveil, F.  1981, Phys. Rev. Lett., 46, 532

\bibitem{hsu} Hsu, J. Y., Matsuda, K., Chu, M. S., $\&$ Jensen, T. H..  1979, Phys. Rev. Lett.,  43, 203

\bibitem{kar} Karney, C. F. F., $\&$   Bears, A. 1977, Phys. Rev. Lett., 39, 550

\bibitem{kar2} Karney, C. F. F.  1978, Phys. Fluids, 21, 1584

\bibitem{kn} Kneer, F., $\&$ Bello Gonz\'{a}lez, N. 2011, A\&A, 532, A111
\bibitem{ono} Ono, M., $\&$  Kulsrud, R.
M. 1975, Phys. Fluids, 18, 1287

\bibitem{sw1} Schwarzschild, M. 1948, ApJ, 107, 1

\bibitem{s}  Sessler, G. M., $\&$  Pearson, G. A. 1967, Phys. Rev., 162, 108


\bibitem{sk1} Smith, G. R., $\&$  Kaufman, A. N. 1975, Phys. Rev. Lett., 34, 1613

\bibitem{sk2} Smith, G. R., $\&$ Kaufman, A. N. 1978, Phys. Fluids, 21, 2230

\bibitem{stix} Stix, T. H.  1992, Waves in  Plasmas (AIP, New York)

\bibitem{v06} Vranjes, J., $\&$ Poedts, S. 2006, A\&A, 458, 635


\bibitem{vs}  Vranjes, J.,  Saleem, H., $\&$ Poedts, S. 2007, Phys. Plasmas, 14, 034504

\bibitem{v08} Vranjes, J., Kono, M.,   Poedts, S., $\&$  Tanaka, M. Y.  2008, Phys. Plasmas, 15, 092107

\bibitem{perm} Vranjes, J.,   Poedts, S., $\&$  Ehsan, Z. 2009,  Phys. Plasmas,  16, 074501
\bibitem{v2} Vranjes, J., $\&$  Poedts, S. 2009a,  Europhys. Lett.,  86, 39001
\bibitem{v3} Vranjes, J., $\&$ Poedts, S. 2009b,  MNRAS, 400, 2147
\bibitem{v09} Vranjes, J., $\&$ Poedts, S. 2009c A\&A, 503, 591

\bibitem{v4} Vranjes, J., $\&$ Poedts, S. 2010a, ApJ,  719, 1335
\bibitem{v5} Vranjes, J., $\&$ Poedts, S. 2010b, MNRAS, 408, 1835

\bibitem{v6} Vranjes, J.  2011a, A\&A,  532, A137
\bibitem{v6a} Vranjes, J.  2011b, MNRAS,  415, 1543
\bibitem{v9} Vranjes, J. 2011c,  Phys. Plasmas,  18, 084501

\bibitem{v11} Vranjes, J.  2011d, Phys. Plasmas, 18, 062902

\bibitem{wah} Wahlberg, C., $\&$    Revenchuk,  S. M. 2003,  Phys. Plasmas,  10, 1164

\bibitem{wei} Weiland, J. 2000, Collective Modes in Inhomogeneous Plasmas (IOP Publishing, Bristol and Philadelphia)


\end{thebibliography}
\end{document}